\documentclass[sigconf]{acmart}
\AtBeginDocument{%
  }

\usepackage{tcolorbox}

\renewcommand\footnotetextcopyrightpermission[1]{}
\begin{document}

\title{LLM-based Semantic Search for Conversational Queries in E-commerce}
\thanks{This is a preprint under review.}

\author{Emad Siddiqui}
\affiliation{%
  \institution{University of Arizona}
  \city{Tucson}
  \state{AZ}
  \country{USA}}
\email{emadsiddiqui@arizona.edu}

\author{Venkatesh Terikuti}
\affiliation{%
  \institution{University of Arizona}
  \city{Tucson}
  \state{AZ}
  \country{USA}}
\email{venkatesht@arizona.edu}

\author{Xuan Lu}
\affiliation{%
  \institution{University of Arizona}
  \city{Tucson}
  \state{AZ}
  \country{USA}}
\email{luxuan@arizona.edu}

\renewcommand{\shortauthors}{Siddiqui et al.}

\begin{abstract}
  Conversational user queries are increasingly challenging traditional e-commerce platforms, whose search systems are typically optimized for keyword-based queries. We present an LLM-based semantic search framework that effectively captures user intent from conversational queries by combining domain-specific embeddings with structured filters. To address the challenge of limited labeled data, we generate synthetic data using LLMs to guide the fine-tuning of two models: an embedding model that positions semantically similar products close together in the representation space, and a generative model for converting natural language queries into structured constraints. By combining similarity-based retrieval with constraint-based filtering, our framework achieves strong precision and recall across various settings compared to baseline approaches on a real-world dataset. %\textcolor{red}{on a modified Amazon ESCI subset}.
\end{abstract}

\begin{CCSXML}
<ccs2012>
   <concept>
       <concept_id>10002951.10003317.10003325</concept_id>
       <concept_desc>Information systems~Information retrieval query processing</concept_desc>
       <concept_significance>500</concept_significance>
       </concept>
   <concept>
       <concept_id>10002951.10003317.10003338</concept_id>
       <concept_desc>Information systems~Retrieval models and ranking</concept_desc>
       <concept_significance>500</concept_significance>
       </concept>
   <concept>
       <concept_id>10010147.10010178.10010179</concept_id>
       <concept_desc>Computing methodologies~Natural language processing</concept_desc>
       <concept_significance>300</concept_significance>
       </concept>
 </ccs2012>
\end{CCSXML}

\ccsdesc[500]{Information systems~Information retrieval query processing}
\ccsdesc[300]{Information systems~Retrieval models and ranking}
\ccsdesc[300]{Computing methodologies~Natural language processing}

\keywords{semantic search, LLM, sentence transformer, synthetic data, e-commerce}


\maketitle

\section{Introduction}
Web users have been trained to search using keywords for decades, particularly since the rise of Google in the early 2000s. However, this is beginning to change with the emergence of large language models (LLMs) and their integration into many aspects of daily life~\cite{Reid2025AIModeSearch}\cite{Bloomreach2025ConversationalAIShoppingStudy}. As users grow accustomed to interacting with LLMs through natural language, their search behavior may shift accordingly—they are increasingly phrasing their queries in natural, conversational language, similar to how they communicate with LLMs. This shift is already evident in recent empirical findings: a 2025 survey shows that 54\% of consumers have shifted toward conversational search habits, favoring natural language over keywords~\cite{Bloomreach2025ConversationalAIShoppingStudy}. E-commerce platforms, which rely heavily on search functionality, are increasingly challenged by this emerging trend. For example, instead of searching for ``\textit{iPhone 14 Pro case}'' and manually applying filters such as review scores, a user may prefer to enter a query that includes all their requirements, such as ``\textit{Find me a iPhone 14 Pro case in green color priced between \$15 and \$20 with strong feedback.}'' However, as e-commerce search systems are typically optimized for keyword-based queries, they often struggle to capture user intent. Research shows that 69\% of consumers use the search bar immediately upon visiting an e-commerce site, yet 80\% abandon the platform due to unsatisfactory search results~\cite{Kashyap23}.

To effectively capture user intent from a search query expressed in conversational natural language, especially with requirements stated explicitly or implicitly, a search system must be able to accurately extract and structure those intents. Additionally, a search system needs to learn and index products in a way that is both effective and efficient for retrieval. Transformers~\cite{vaswani2017attention} have been widely adopted for such tasks, with existing work primarily based on BERT-like models that focus on word-level representations. In recent years, Sentence Transformers~\cite{reimers2019sentence} have gained increasing attention for their ability to capture semantic meaning at the sentence level, making them more suitable for this new setting~\cite{Aamir2024}.

Despite recent advances, two main challenges remain in achieving effective semantic search for e-commerce. First, Sentence Transformers need to be fine-tuned to adapt to specific domains, but labeled data is often scarce, making it unclear which products should be positioned closer together in the embedding space. Second, embeddings often struggle to represent numerical values and categorical information~\cite{wallace2019do}, which may encode key requirements, as illustrated in the example query above.

To address these challenges, we present a LLM-based semantic search system that effectively captures user intent from conversational search queries and retrieves relevant products from large-scale catalogs.  
Specifically, we propose generating synthetic queries for products by leveraging LLMs' world knowledge and reasoning capabilities, and using the inherent semantic links between these synthetic queries and product information to guide the fine-tuning of Sentence Transformers. The same fine-tuned Sentence Transformer is then used to embed user queries for similarity-based retrieval. 
To capture numerical values and categorical information, we fine-tune a generative model to convert user input into structured filters that are applied before retrieval, ensuring that only items meeting the extracted constraints are considered. To achieve this, we continue using an LLM to enrich the synthetic queries with numerical and categorical constraints and generate corresponding structured filters, which are then used to guide the fine-tuning of the selected generative model.

We use a large-scale Amazon Review dataset~\cite{hou2024bridging}, covering 1.3 million products, to fine-tune our models. To evaluate the semantic search performance of the proposed system, we construct a test set based on query–product relevance annotations from the Amazon ESCI dataset~\cite{reddy2022shopping}. 
Experimental results show that our system achieves precision@$k$ scores of 0.32, 0.20, and 0.13 for $k=1, 5, 10$, respectively, and recall@$k$ scores of 0.16, 0.44, and 0.57 for $k=1, 5, 10$, respectively, significantly outperforming baselines.

Our contributions can be summarized as follows:

\begin{itemize}
    \item We propose a framework to investigate how to capture user intent from conversational queries with requirements, expressed either explicitly or implicitly, to improve e-commerce search performance. Our approach, which combines domain-specific embeddings with structured filters, achieves strong precision and recall compared to baseline methods.
    \item We propose using LLMs to generate synthetic data to guide domain-specific fine-tuning of models at various stages. The generated queries (both raw and enriched), along with their corresponding structured filters, convey inherent semantic links to products, thereby contributing to effective semantic search.
    \item We create a set of user queries that reflect diverse levels of complexity in user requirements, along with carefully generated labels within the Amazon ESCI dataset. We release this dataset as a benchmark to support future research on conversational search queries.\footnote{\url{https://anonymous.4open.science/r/Conversational-Search-Query-Benchmark-Based-on-ESCI-S-06DB/}}
\end{itemize}

\section{Related work}

We briefly discuss related work in the following areas within the scope of e-commerce.

\noindent $\bullet$ \textbf{Conversational Search.}
Conversational search~\cite{radlinski2017theoretical, yang2018query} explores how to support natural language interactions between users and information retrieval systems in a dialogue setting, with the goal of collecting user needs in a natural and efficient manner. For example, prior research has investigated a multi-turn dialogue approach~\cite{liu2020enhancing} that incorporates attention mechanisms to improve customer service interactions. A related task, known as conversational recommendation~\cite{liu2023u, jia2022convrec}, aims to recommend items to users based on information extracted from dialogues. Our focus differs in that we target search queries rather than full dialogues in e-commerce, referred to as \textit{conversational queries}, which are single-turn inputs where user intent is fully expressed.

\noindent $\bullet$ \textbf{Semantic Search.}
Semantic search aims to address the limitations of keyword-based retrieval by learning dense representations of queries and items. Transformers~\cite{vaswani2017attention} (such as ERT-like models~\cite{DBLP:journals/corr/abs-1810-04805}), including Sentence Transformers~\cite{reimers2019sentence} (such as Sentence-BERT~\cite{reimers-2019-sentence-bert}), have been widely adopted in various retrieval tasks, including e-commerce. However, existing work typically uses pre-trained transformers or fine-tunes them on domain-specific data. For example, Nigam et al.\cite{Nigam19} demonstrated that sentence-level embeddings trained on click data can significantly improve relevance across multi-million–item catalogs. Our work goes further by introducing synthetic queries to guide transformer fine-tuning, leveraging the rich knowledge and reasoning capabilities of LLMs.

\noindent $\bullet$ \textbf{Synthetic Data.}
Cho et al.~\cite{cho-etal-2022-query} introduced the idea of using pre-trained language models to generate synthetic queries for every document, enabling dense retrievers to be trained without manual annotations. Extensions of this approach to e-commerce include the work of Chaudhary et al.\cite{Chaudhary23}, who evaluated synthetic queries on three public retail datasets, and Jagatap et al.\cite{Jagatap2024}, who addressed cold-start ranking in new categories. However, these studies focus exclusively on retrieval or ranking. Our work uses synthetic data at various stages of the semantic search system, including not only the similarity-based retrieval but also the structured filter extraction. 

\noindent $\bullet$ \textbf{Structured Filter Generation.}
Extracting structured constraints from search queries (e.g., price ranges, categories, or review scores) is crucial for satisfying user requirements and is complementary to embedding-based methods. Traditional approaches primarily rely on rule-based techniques. Loughnane et al.\cite{loughnane-etal-2024-explicit} proposed a transformer-based NER pipeline for identifying spans like ``price'' or ``color.'' While their pipeline focuses on span detection and relies on rule-based logic for filter application, we take a different approach by training a generative model to directly produce structured filters from natural language queries. 
Toolkits such as LangChain’s self-querying retriever~\cite{LangChainSelfQuery} demonstrate how an LLM can generate structured filters that are applied after an similarity-based search. While promising, these demonstrations have been limited to small-scale document collections and do not address the scalability, and catalog complexity challenges inherent in retail search systems operating at the scale of millions of items.

\section{Dataset}

We used the Amazon Reviews 2023 dataset~\cite{hou2024bridging}, focusing on the Cell Phones \& Accessories category. The dataset contains approximately 1.3 million products spanning the years 1996 to 2023. Each product record includes a unique identifier, textual information including \textit{title of the product}, \textit{description}, \textit{features}, and \textit{technical specifications}, numerical information including \textit{price}, the \textit{number of reviews}, and the \textit{average rating} of the product, and a label of \textit{subcategory}. 

\noindent $\bullet$ \textbf{Dataset Preprocessing.}
We apply a two-step process to preprocess the dataset. First, the textual information was cleaned by (1) removing HTML tags, URLs, and non-ASCII characters; (2) converting all text to lowercase; and (3) trimming whitespace. Second, since a manual check revealed instances of incorrect product categorization, we corrected the subcategory labels using the Gemini Flash reasoning model~\cite{google_gemini_api_models_2025}, which is scalable to large datasets like ours through API usage. Specifically, given a product’s textual information, the model was instructed to assign the product to either the \textit{Cell Phones} or \textit{Cell Phone Accessories} subcategory. Discrepancies between the newly assigned labels and the original labels are reviewed to ensure correctness, contributing to improved reliability for downstream tasks such as filtering. 

\begin{figure*}
    \centering
    \includegraphics[width=0.9\linewidth]{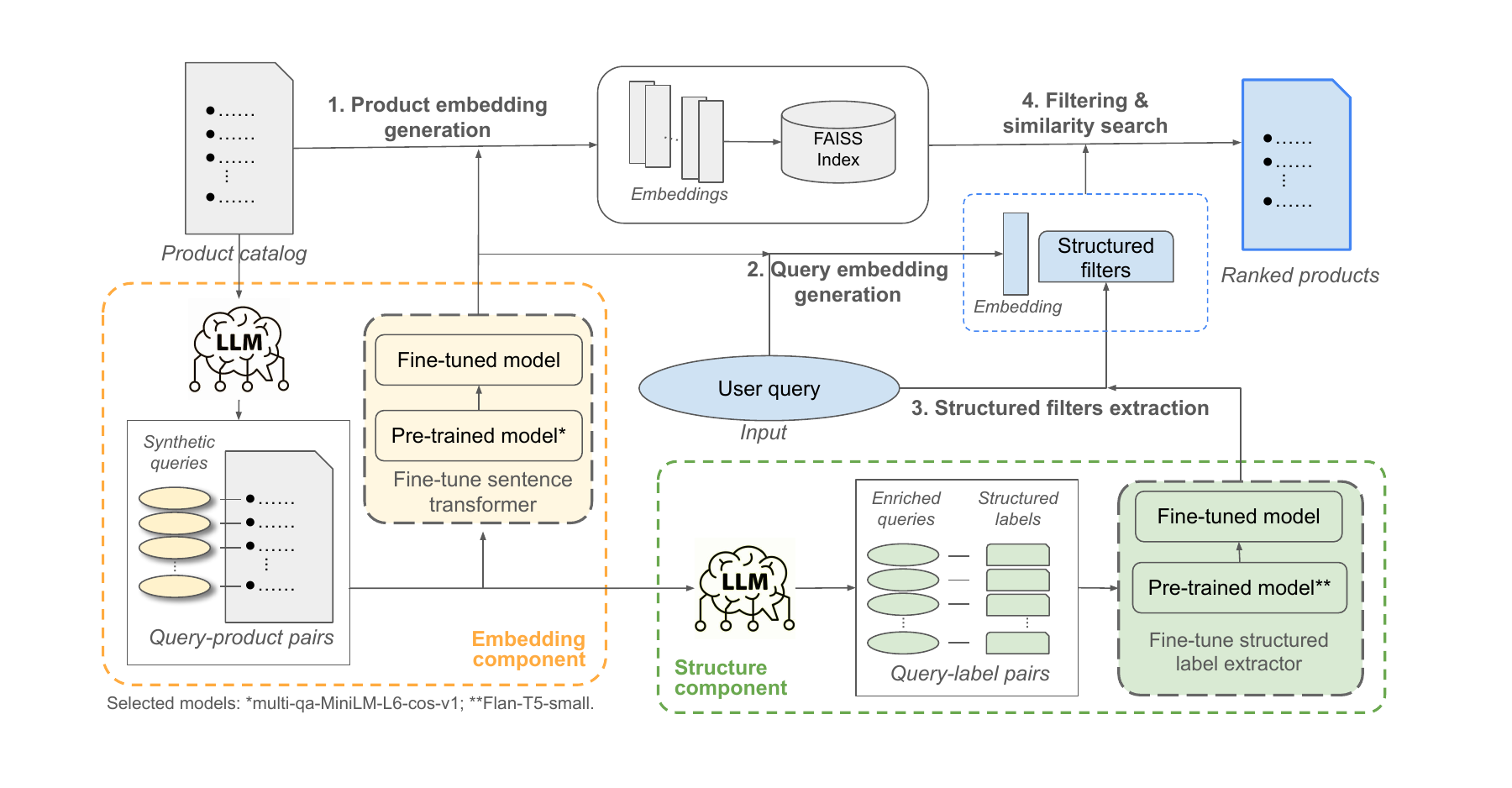}
    \caption{The LLM-based Semantic Search Framework for Conversational Queries. For each input user query, the framework outputs a ranked list of relevant products by combining similarity search with constraint-based filtering. \textit{Embedding component}: this component fine-tunes a Sentence Transformer using synthetic queries generated by an LLM from the product catalog. The fine-tuned model is then used to compute embeddings for both the products and the user query. \textit{Structure component}: this component fine-tunes a generative model to extract numerical and categorical information from user queries and convert it into structured filters. Note that the target product catalog used to generate product embeddings may differ from the catalog used to train the embedding component.}
    \label{fig:framework}
\end{figure*}

\section{Methods}

As shown in Figure~\ref{fig:framework}, our framework computes semantic embeddings for each product in the product catalog.  
For each user query, the system extracts user intent by combining an embedding with structured filters. It then excludes irrelevant products using the structured filters, and finally produces a ranked list of products based on similarity scores using the query embedding. We now describe our methods for generating meaningful embeddings and accurate structured filters, corresponding to the embedding and structure components in Figure~\ref{fig:framework}, respectively. 
All training was conducted on a single NVIDIA V100 GPU (32GB), which provids sufficient memory for efficient fine-tuning of both models.

\subsection{Product Embedding Generation}
A key component of our framework is generating embeddings for all products in the target catalog to enable accurate and efficient retrieval for user queries. To achieve this, we fine-tune a sentence transformer using synthetic queries to fully capture semantic information in the embeddings and use FAISS to index them for accelerated retrieval.

\subsubsection{Sentence transformer selection.} Sentence Transformers is a widely used framework for generating dense vector representations that captures the overall meaning of an entire sentence or paragraph unlike standard transformer models, which produce token-level embeddings. This sentence-level representation enables efficient similarity computation for downstream tasks like semantic search and retrieval~\cite{reimers-2019-sentence-bert}, suitable for applications requiring semantic similarity. We select \texttt{multi-qa-MiniLM-L6-cos-v1}~\cite{multiqaMinilm2024}, a model that well balances accuracy and latency with a lightweight architecture that can be efficiently scaled to millions of records without compromising retrieval quality. Additionally, this model offers an extended input capacity (i.e., 512 tokens), which is beneficial for long texts such as ours (i.e., product title, description, features, and technical specifications).

\subsubsection{Synthetic query generation.} \label{subsubsec:synthetic}
While using a pre-trained sentence transformer is the default option for generating embeddings, fine-tuning the model with domain-specific information typically improves performance. In addition, we aim for the embedding algorithm to learn how to represent product information in a way that aligns with the user intent expressed in potential queries for that product—so that the product is identified as a correct match when such a query occurs.

\begin{figure}[t]
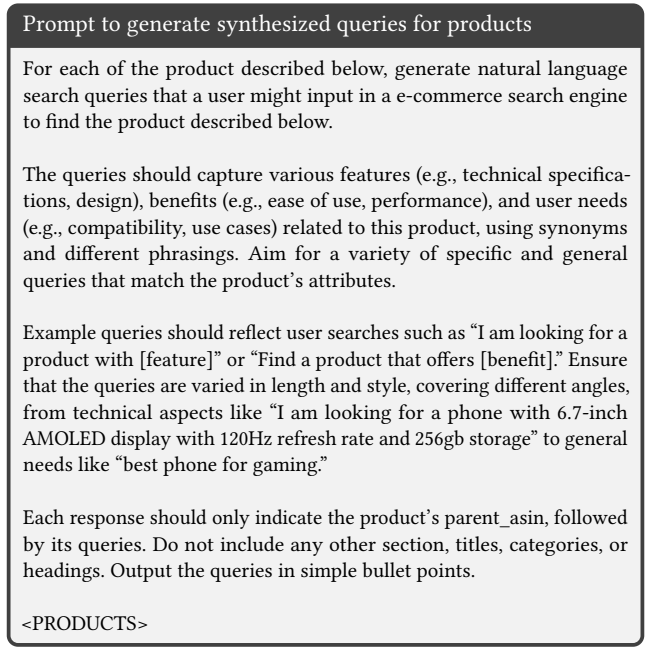

    \centering
        \begin{tcolorbox}[
        left=2pt, right=2pt, top=2pt, bottom=2pt,
        fontupper=\small,
        title=Prompt to generate synthesized queries for products]
    For each of the product described below, generate natural language search queries that a user might input in a e-commerce search engine to find the product described below. \\
    
    The queries should capture various features (e.g., technical specifications, design), benefits (e.g., ease of use, performance), and user needs (e.g., compatibility, use cases) related to this product, using synonyms and different phrasings. Aim for a variety of specific and general queries that match the product's attributes. \\
    
    Example queries should reflect user searches such as ``I am looking for a product with [feature]'' or ``Find a product that offers [benefit].'' Ensure that the queries are varied in length and style, covering different angles, from technical aspects like ``I am looking for a phone with 6.7-inch AMOLED display with 120Hz refresh rate and 256gb storage'' to general needs like ``best phone for gaming.'' \\
    
    Each response should only indicate the product's parent\_asin, followed by its queries. Do not include any other section, titles, categories, or headings. Output the queries in simple bullet points.\\
 \\
    <PRODUCTS>
        \end{tcolorbox}
    \caption{Prompt to generate synthetic queries for products. The <PRODUCTS> placeholder is replaced with a list of products, each described by its unique identifier (i.e., parent\_asin), product title, features, description, and technical specifications.}
    \label{fig:query_prompt}
\end{figure}

To achieve this, we generate query-product pairs, where the queries are synthesized from products' textual information, which merges product title, description, features, and technical specifications, using a Gemini Flash model. The Gemini models (used here and in other parts of this work) are selected for their API accessibility, large context window (which allows processing multiple products in a single prompt, thereby improving efficiency), and free-tier availability. Other LLMs including open-weight models may also be capable of performing this task. We employ zero-shot prompting to generate high-quality synthetic queries. As shown in Figure~\ref{fig:query_prompt}, the model is instructed to return multiple synthetic queries for each product. We apply this prompt to half of the products in our dataset, with each product receiving approximately 10 diverse synthetic queries, resulting in a total of 6,503,773 unique query-product pairs. This covers 669,294 unique products and 3,452,309 unique queries. 

We randomly sample 100 products (50 cell phones and 50 accessories) together with their corresponding synthetic queries to demonstrate that the generated queries are relevant, diverse, and natural-sounding, and  closely resemble real user behaviors. We provide an analysis of the generated queries with representative examples in the Appendix. 

\subsubsection{Model fine-tuning.} We then use the 6.5 million query-product pairs to fine-tune the sentence transformer. Each training sample consists of a query and the textual information of the corresponding product. We use \texttt{MultipleNegativesRankingLoss}~\cite{SbertLosses} as the loss function defined as follows:

\begin{equation}\label{eq:loss}
    \mathcal{J}(\mathbf{x}, \mathbf{y}, \theta) = -\frac{1}{K} \sum_{i=1}^{K} \left[ S(x_i, y_i) - \log \sum_{j=1}^{K} e^{S(x_i, y_j)} \right]
\end{equation}

where \( \mathcal{J}(\mathbf{x}, \mathbf{y}, \theta) \) denotes the approximated mean negative log-likelihood of the correct responses in a batch of items \( \mathbf{x} = (x_1, x_2, \dots, x_K) \) (i.e., synthetic queries) and corresponding correct responses \( \mathbf{y} = (y_1, y_2, \dots, y_K) \) (i.e., products), with a batch size \(K\) set to 168 in our case. Recall that the synthetic queries are generated using an LLM based on product information, the product paired with a query is regarded as the query's correct response. The similarity function $S$, parameterized by the neural network weights $\theta$, is used to measure how well a query matches a product. The model is trained to minimize $\mathcal{J}$, thereby assigning higher similarity scores to generated query-product pairs (i.e., $(x_i, y_i)$) to mismatched pairs (i.e., $(x_i, y_j)$ where $i \ne j$), effectively learning to rank relevant responses higher~\cite{Henderson17}. For each batch, we ensure that every selected product or synthetic query appears only once to avoid false negatives. As a result, a product is more likely to be retrieved and ranked highly when a query similar to the one paired with that product is issued. 

\subsubsection{Indexing embeddings}

The fine-tuned sentence transformer is then used to compute embeddings for all products in the catalog for application. To enable fast and scalable similarity search across large product catalog, we utilize Facebook AI Similarity Search (FAISS)~\cite{douze2024faiss} to index the embeddings. FAISS is a highly efficient library for searching and clustering dense vector representations.  
Furthermore, to ensure that only products meeting the structured constraints are considered during retrieval, we develop an IVF-Flat index~\cite{faiss_wiki_indexes} for product embeddings, where all products are stored according to their metadata. During retrieval, structured constraints extracted from the query are applied using FAISS’s ID filtering mechanism. 

Specifically, the IVF-Flat index partitions the embedding space into a set of inverted lists, with the number of partitions determined heuristically based on the corpus size. After training the index from a random sample of product embeddings, all product embeddings are added to it along with deterministic integer IDs corresponding to their row indices in the product metadata table. This one-to-one mapping ensures stable alignment between each FAISS vector and its associated metadata.  
For retrieval, with the structured constraints extracted from the query, we preselect the set of product IDs whose metadata satisfy these constraints, and pass them to FAISS to restrict the search space to only the matching subset of vectors. FAISS then performs similarity search (inner product) between the query embedding and the filtered vectors, retrieving the top-$k$ most relevant products. This two-stage approach combines the efficiency of FAISS’s partitioned search with exact filtering over structured attributes, ensuring that retrieved results are both semantically relevant and constraint-compliant.

\subsection{User Query Understanding}
Using the same fine-tuned sentence transformer, we can learn an embedding for each user query, and retrieve relevant products based on similarity scores.  However, embeddings often struggle to represent numerical values and categorical information accurately~\cite{wallace2019do}\cite{kim-etal-2019-categorical}. To fully capture user intent from their queries, we choose to combine the embedding with structured filters~\cite{LangChainSelfQuery}. Specifically, 
we fine-tune a language model, \texttt{Flan-T5-small}~\cite{flan-t5-small}, to extract structured criteria from the numerical and categorical information of the queries. \texttt{Flan-T5-small}'s instruction-tuned design ensures consistent and structured outputs. 
The criteria include price, rating, number of reviews, and the subcategory of the target product, as these dimensions are available in the structured attributes of the review dataset. These criteria are used to filter the products before the retrieval based on similarity scores. Final ordering is then determined by dense retrieval scores. 

\subsubsection{Structured filter generation for synthetic queries}\label{subsubsec:label_ge}
Using the synthetic queries generated in Section~\ref{subsubsec:synthetic}, we further employ the Gemini Flash model to enrich the queries with preferences related to one or more of the criteria dimensions mentioned above, and extract and structure the corresponding filters. Table~\ref{tab:label_example} shows an example. We further employ a chat-based generative model (ChatGPT) to synthesize queries that target certain underrepresented constraint styles, such as cell phones with both minimum and maximum price limits, thereby increasing coverage and lexical diversity in the constraint space.

\begin{table}[h!]
\centering
\small
\caption{A synthetic query is enriched with numerical features, and structured filters are extracted and formatted in \texttt{JSON}.}
 \begin{tabular}{l}
\midrule
Original query: \\
\begin{minipage}{.46\textwidth}
\begin{verbatim}

smartphone with good battery life

\end{verbatim}
\end{minipage} \\
\hline
Enriched query: \\
\begin{minipage}{.46\textwidth}
\begin{verbatim}

smartphone with good battery life, plenty of reviews and 
priced under $300

\end{verbatim}
\end{minipage} \\
\hline
Generated label: \\
\begin{minipage}{.46\textwidth}
\begin{verbatim}

"price": {"min": null, "max": 300}, 
"rating_number": {"min": "high", "max": null}, 
"average_rating": null

\end{verbatim} 
\end{minipage}\\
\hline
Structured filters: \\
\begin{minipage}{.46\textwidth}
\begin{verbatim}

{"price_min": null, "price_max": 300.0, 
"review_count_min": "high", "review_count_max": null, 
"average_rating_min": null, "average_rating_max": null, 
"subcategory": "Cell Phones"}

\end{verbatim} 
\end{minipage}\\
\midrule
\end{tabular}

 \label{tab:label_example}
\end{table}

The constraints can be quantitative or qualitative, depending on the description in the enriched query. For numerical values, such as \textit{``priced under \$300''} in the example, the model sets explicit numerical thresholds, i.e., setting \texttt{price\_max} to be $300$. For qualitative descriptions, such as \textit{``plenty of reviews''}, the model maps such expressions to standardized values of \texttt{high}, \texttt{medium}, or \texttt{low}. In the example, the label of \texttt{review\_count\_min} is set to be \texttt{high}. To map the values of \texttt{high}, \texttt{medium}, and \texttt{low} to the product information when searching, we adopt thresholds listed in Table~\ref{tab:threshold}. 
We map qualitative filters to numeric thresholds based on user expectations. This ensures that qualitative terms such as cheap, popular, or highly rated are applied consistently during retrieval, while also allowing flexibility to adjust these thresholds as needed. Note that these mappings are application-dependent and can be tuned per catalog, and our system supports dynamic threshold selection. In our experiments, however, we fix the thresholds to support training of the single-vector baseline and to ensure that different approaches are evaluated against the same targets. The thresholds used in this work were chosen based on experience and may not be optimal. 

\begin{table}[h]
\centering
\caption{Thresholds for rating, review count, and price levels.} 
\begin{tabular}{lccc}
\toprule
\textbf{Metric} & \textbf{Low} & \textbf{Medium} & \textbf{High} \\
\midrule
Rating   &  [0, 4.0)         & [4.0, 5]       & [4.5, 5] \\
\#Reviews & [0, 100)         & [100, +$\infty$)       & [1000, +$\infty$)\\
Price &  &  &  \\
\hspace{1em}\textit{Cell Phones} & [0, 100] & [100, 300] & [300, +$\infty$) \\
\hspace{1em}\textit{Cell Phone Accessories} & [0, 15] & [15, 40] & [40, +$\infty$) \\
\bottomrule
\end{tabular}
\label{tab:threshold}
\end{table}

The resulting dataset contains 61,812 query–filters pairs. We randomly select 1,000 pairs for evaluation and filters in this set are manually validated and corrected as needed. The remaining pairs of this set is used for model fine-tuning.

\subsubsection{Model fine-tuning} \label{subsubsec:structure_fine_tune}
To enable accurate extraction of structured filters, we adopt a sequence-to-sequence setup to generate a structured text output conditioned on the enriched input query. This formulation allows the model to learn how to map unstructured user intents to standardized filter representations. Specifically, we fine-tune the \texttt{Flan-T5-small} model using the enriched query-label pairs with a \texttt{Seq2SeqTrainer}~\cite{HuggingFaceTrainer}, minimizing token-level cross-entropy loss over the target sequence~\cite{jurafsky2025slp3}. At each decoding step \( t \), the loss is computed based on the log-probability assigned to the correct next token \( w_{t+1} \), as follows:

\begin{equation}
\mathcal{L}_{\text{CE}} = - \log \hat{y}_t[w_{t+1}] 
\end{equation}

where \( \hat{y}_t \) is the model’s predicted distribution over the vocabulary and \( w_{t+1} \) is the correct next token. The total loss is averaged over all tokens in the target output. Training is performed using teacher forcing, where the model is conditioned on the correct sequence history rather than its own previous predictions \cite{jurafsky2025slp3}.

\section{Evaluation}
We then evaluate our framework and report its performance. First, we assess the user intent extraction component based on its accuracy in generating structured filters for queries. Then, we evaluate the entire framework based on its ability to retrieve relevant products.

\subsection{Test Setting}

\begin{table}
\small
  \caption{Sample Test Queries.}
  \label{tab:sample_test_queries}
  \begin{tabular}{cp{7cm}}  
    \toprule
    \textbf{Query \#} & \textbf{Natural Language Query} \\
    \midrule
    1 & 4G basic phones with keyboards  \\
    2 & AT\&T prepaid phones under \$200 with 4+ stars. \\
    3 & Huawei P30 Pro unlocked. Maximum price: \$300. \\
    4 & GSM unlocked flip phones with strong customer feedback \\
    5 & Show me 6-inch screen phones between \$100 and \$200 and rated 4.2+ stars from 250+ reviews. \\
    6 & Show me Alice in Wonderland iPhone 7 Plus cases with decent review count. \\
    7 & Anker 4-port USB charger averagely priced \\
    8 & I'm searching for a slim waterproof 40 mm Apple Watch Series 4 band with a regular buckle under \$25 with strong ratings. \\
    9 & I'm looking for an athletic phone holder between \$10 and \$14. \\
    10 & I need a cheap and big iPhone SE case. \\
    \bottomrule
  \end{tabular}
\end{table}

Annotated relationships between queries and products is essential for evaluating a system's ability to answer queries with high precision and recall. However, manually annotating whether a given product is a positive match for every query is infeasible when the product catalog contains 1.3 million items.
To address this challenge, we leverage the Amazon ESCI (Shopping Queries) dataset~\cite{reddy2022shopping}, which provides query-product relevance annotations with 130,652 unique queries and 2.62 million query–product judgments. From its extension version, ESCI-S, which augments the original product catalog with structured metadata (such as price,\footnote{For products lacking price information, we generate plausible price values using the \texttt{Gemma 3:12B} local LLM, conditioning on the available product information.} average rating, and review count) for approximately 1.66 million products, we select roughly 22,000 products belonging to the \textit{Cell Phones \& Accessories} category. This catalog serves as the testbed for our framework, on which we evaluate the performance of our proposed system and the baselines in retrieving matched products for given queries.

To create the test queries, for the 22,000 products, we extract query–product pairs that include these products and are labeled as exact matches in the original ESCI dataset. We then randomly sample 151 queries from this set and rewrite them into natural language queries enriched with additional constraints. These constraints deliberately narrow the set of products that satisfy each query, creating a more challenging evaluation set and enabling systematic assessment of our system’s precision and recall relative to the baseline model. 

Most of the 151 queries are complex and attribute-rich to reflect realistic user intent involving multiple constraints (such as device type, specific technical specifications, and price). A few simpler queries, such as “Apple 11 Pro Max,” are included to assess the framework’s robustness in handling varying levels of query complexity. All queries are phrased in natural language and structured to require semantic understanding for effective retrieval. See examples in Table~\ref{tab:sample_test_queries}. Of these queries, 50 have exactly one match, 87\% have up to four matches, and none has more than eight exact matches in the catalog of approximately 22,000 products. Therefore, we restrict our evaluation of top-$k$ retrieval effectiveness to values where $k$ $\leq$ 10. Metrics at $k$ = 10 evaluate the system’s ability to retrieve all relevant items, whereas metrics at smaller $k$ values (e.g., $k$ $\leq$ 5) are more indicative of ranking quality and thus more directly affect the user search experience.

\subsection{Baselines}

We compare our system with the following baselines.

\noindent $\bullet$ \textbf{\textit{Pre-trained Sentence Transformer}}. In this setup, we use the pre-trained \texttt{multi-qa-MiniLM-L6-cos-v1} model from the Sentence Transformers library without any task-specific fine-tuning.  
To introduce basic filter awareness, we append price, rating, and review count information directly to the beginning of the product metadata prior to embedding. Placing these attributes at the start helps ensure they are encoded as part of the product representation.
    
\noindent $\bullet$ \textbf{\textit{Pre-trained Sentence Transformer + structured filter extraction}}. In this setup, we combine the same pre-trained Sentence Transformer with our fine-tuned Flan-T5-small filter extraction model. We extracted structured constraints directly from the query using Flan-T5-small and applied them as pre-filters to the retrieved results.

\noindent $\bullet$ \textbf{\textit{Sentence Transformer fine-tuned with prepended metadata}}. This baseline uses the same sentence transformer architecture as our system but is fine-tuned differently. Specifically, price, rating, and review information are prepended to the product text during training, and the model embeds the full product description together with this metadata to provide a fair single-vector baseline for comparison with our multi-stage retrieval approach. While our system’s transformer is trained on approximately 6.5 million query–product pairs without explicit price, rating, or review constraints, this baseline is fine-tuned on a modified dataset that incorporates such constraints. Starting from the original dataset, we first remove around 487 thousand pairs containing qualitative keywords such as best, premium, and top-rated, which implicitly correspond to price, rating or review filters. We then add approximately 780 thousand additional query–product pairs in which the queries contain explicit price, rating, or review count constraints, and the paired products satisfy those constraints.
To construct this augmented dataset, we began with roughly 1 million enriched queries generated using the Gemini API. We applied our filter extraction model to identify filter constraints (price, rating, and review count) in each query and separated pairs where product metadata met those constraints. For the remaining pairs, we retained only queries containing purely numeric constraints and programmatically replaced their constraint values with randomized but valid alternatives that matched the paired product metadata. This procedure increased the diversity of constraint formulations. We did not modify queries containing qualitative constraints (e.g., cheap, expensive), since these are mapped to “low,” “medium,” or “high” ranges by our extractor and the original string values cannot be deterministically recovered.

\noindent $\bullet$ \textbf{\textit{Sentence Transformer fine-tuned with prepended metadata + structured filter extraction}}. In this baseline, we combine the fine-tuned Sentence Transformer with our fine-tuned Flan-T5-small filter extraction model similar to baseline 2.

\noindent $\bullet$ \textbf{\textit{Sentence Transformer fine-tuned without prepended metadata}}. We also provide a baseline using the fine-tuned sentence transformer, identical to our complete system except for the absence of the structured filtering component. 

For structured filter extraction, we do not include alternative models in the baseline settings but proceed directly with the fine-tuned \texttt{Flan-T5-small} model due to its strong performance, which can be directly evaluated using our held-out query-label dataset. We explicitly compare it with a NER model in Section~\ref{subsub:label_extract_eva}, but readers may skip to the results in Section~\ref{sub:results} if they are not interested in the details.

\subsubsection{Label extraction component evaluation}\label{subsub:label_extract_eva}

With the held-out set of 1,000 query-label pairs generated in Section~\ref{subsubsec:label_ge}, we are able to directly evaluate the label extraction component. Across all label dimensions (such as \textit{price\_min} and \textit{subcategory}, see the label structure in Table~\ref{tab:label_example}) the fine-tuned \texttt{Flan-T5-Small} achieves strong results with an accuracy of 99.8\% - 99.9\%, and the overall match accuracy reaches 99.4\%. 

The high customizability of the \texttt{Flan-T5-Small} model may help explain its strong performance in this task. For example, it performs well when fine-tuned to map subjective phrases such as “great reviews” to a combination of \texttt{average\_rating\_min = high} and \texttt{review\_count\_min = medium}. Alternatively, the same phrase can be trained to correspond solely to \texttt{average\_rating\_min = high}, depending on the desired level of specificity. Similarly, phrases like “best” can be flexibly mapped to different filter combinations based on application needs.

We introduce a baseline to further demonstrate that the exceptionally high performance likely stems from the capabilities of the language model, rather than issues such as data bias.

\noindent $\bullet$ \textbf{\textit{BERT-based NER Baseline}}. 
We use a rule-driven Named Entity Recognition (NER) baseline to compare its structured filter extraction ability with our LM-based method. We choose the \texttt{bert-base-NER} model~\cite{DBLP:journals/corr/abs-1810-04805}, a widely used BERT-based model fine-tuned on NER datasets, known for its strong out-of-the-box performance and ease of integration. To adapt it to our use case, we wrap the NER pipeline with domain-specific, rule-based logic. Specifically, for the product subcategory (i.e., \textit{Cell Phones} or \textit{Cell Phone Accessories}), we use a keyword-based strategy to determine it with a predefined list of accessory-related terms (such as ``charger'' and ``screen protector''). For numerical information, related entities are  interpreted using context-aware heuristics. For instance, if a numeric span is preceded by phrases like ``under,'' ``less than,'' or ``stars,'' we infer a corresponding constraint on price\_max or average\_rating\_max. Additionally, we integrate qualitative cues using handcrafted keyword lists. Phrases such as “highly rated,” “cheap,” or “many reviews” are mapped to categorical thresholds (e.g., average\_rating\_min = high, price\_max = low, review\_count\_min = high).

\begin{table}
  \caption{Comparison of Label Extraction Accuracy (\%) Between NER Baseline and Fine-Tuned Flan-T5-Small.}
  \label{tab:filter_extraction_accuracy_comparison}
  \begin{tabular}{lcc}
    \toprule
    \textbf{Field} & \textbf{NER Baseline} & \textbf{Flan-T5-Small} \\
    \midrule
    price\_min            & 85.60  & 99.9 \\
    price\_max            & 59.10  & 99.8 \\
    review\_count\_min    & 68.60  & 99.8 \\
    review\_count\_max    & 98.80  & 99.8 \\
    average\_rating\_min  & 70.70  & 99.8 \\
    average\_rating\_max  & 100.00 & 99.8 \\
    subcategory           & 84.70  & 99.9 \\
    \midrule
    \textbf{Overall exact match} & 24.40 & \textbf{99.4} \\
    \bottomrule
  \end{tabular}
\end{table}

Table~\ref{tab:filter_extraction_accuracy_comparison} shows the effectiveness of the NER model in extracting structured filters. Our model outperforms it across all dimensions except \texttt{average\_rating\_max}. Notably, only 24.4\% of queries have all label dimensions correctly extracted. Inaccuracies in extracting these structured filters, which are used to filter out irrelevant products, can severely degrade overall retrieval performance. While such results could potentially be improved through keyword expansion or additional post-processing logic, these approaches introduce complexity and may still lack robustness in real-world deployment. We will use the fine-tuned \texttt{Flan-T5-Small} as the label extraction component in the system evaluation.

\subsection{Results}\label{sub:results}

The evaluation results are reported in Table~\ref{tab:retrieval_metrics_combined}. Overall, our system outperforms the baselines  with substantial gains across all evaluation metrics. 

\begin{table*}
\centering
\caption{Evaluation Results Measured by Average Precision and Recall at Top-$k$. When structured filtering is included (bottom table) or not (top table), the best results at each $k$ are shown in bold.}
\label{tab:retrieval_metrics_combined}
\begin{tabular}{ccccccc}
\toprule
& \multicolumn{2}{c}{\begin{tabular}[c]{@{}c@{}}Pre-trained Sentence \\ Transformer (filters prepended)\end{tabular}} & \multicolumn{2}{c}{\begin{tabular}[c]{@{}c@{}}Fine-tuned Sentence \\ Transformer (filters prepended)\\ \end{tabular}} & \multicolumn{2}{c}{\begin{tabular}[c]{@{}c@{}}Fine-tuned Sentence \\ Transformer (No filters prepended)\end{tabular}} \\\hline\hline
k & Precision@k & Recall@k & Precision@k  & Recall@k  & Precision@k & Recall@k  \\
\midrule
1   & 0.11 & 0.06 & \textbf{0.14} & 0.08 & 0.13 & \textbf{0.08} \\
2   & 0.08 & 0.09 & \textbf{0.13} & \textbf{0.14} & 0.12 & 0.13 \\
3   & 0.08 & 0.13 & \textbf{0.12} & \textbf{0.19} & 0.11 & \textbf{0.19} \\
5   & 0.08 & 0.20 & \textbf{0.10} & \textbf{0.25} & \textbf{0.10} & 0.24 \\
10  & 0.05 & 0.26 & 0.07 & \textbf{0.36} & \textbf{0.08} & \textbf{0.36} \\
\bottomrule
\toprule
& \multicolumn{2}{c}{\begin{tabular}[c]{@{}c@{}}Pre-trained Sentence \\ Transformer (filters prepended) \\  \& Fine-tuned Flan-T5-Small\end{tabular}} & \multicolumn{2}{c}{\begin{tabular}[c]{@{}c@{}}Fine-tuned Sentence \\ Transformer (filters prepended) \\  \& Fine-tuned Flan-T5-Small\end{tabular}} & \multicolumn{2}{c}{\begin{tabular}[c]{@{}c@{}}Fine-tuned Sentence \\ Transformer (No filters prepended) \\  \& Fine-tuned Flan-T5-Small (ours)\end{tabular}} \\\hline\hline
k & Precision@k & Recall@k & Precision@k  & Recall@k  & Precision@k & Recall@k  \\
\midrule
1   & 0.22 & 0.11 & 0.30 & \textbf{0.16} & \textbf{0.32} & \textbf{0.16} \\
2   & 0.19 & 0.18 & \textbf{0.29} & \textbf{0.30} & \textbf{0.29} & 0.28 \\
3   & 0.17 & 0.23 & \textbf{0.25} & \textbf{0.37} & \textbf{0.25} & 0.36 \\
5   & 0.14 & 0.32 & \textbf{0.20} & \textbf{0.47} & \textbf{0.20} & 0.44 \\
10  & 0.09 & 0.39 & \textbf{0.13} & \textbf{0.57} & \textbf{0.13} & \textbf{0.57} \\
\bottomrule
\end{tabular}
\end{table*}

\subsubsection{Effectiveness of structured filter extraction}\label{subsubsec:structure_effectiveness}
Adding structured filter extraction (Flan-T5-small) consistently improves both precision and recall across all sentence transformer configurations (compare the top and bottom tables). For the pre-trained sentence transformer with prepended metadata, precision increases from 0.11 to 0.22 at $k{=}1$ and from 0.05 to 0.09 at $k{=}10$, while recall rises from 0.06 to 0.11 ($k{=}1$) and from 0.26 to 0.39 ($k{=}10$). When the model is fine-tuned with prepended metadata, precision improves from 0.14 to 0.30 at $k{=}1$ and from 0.07 to 0.13 at $k{=}10$, with recall gains from 0.08 to 0.16 ($k{=}1$) and from 0.36 to 0.57 ($k{=}10$). A similar pattern holds for the fine-tuned model without prepended metadata: precision rises from 0.13 to 0.32 at $k{=}1$ and from 0.08 to 0.13 at $k{=}10$, and recall improves from 0.08 to 0.16 ($k{=}1$) and from 0.36 to 0.57 ($k{=}10$). These gains indicate that accurate, query-specific pre-filtering substantially reduces irrelevance prior to ranking, yielding the largest absolute improvements in recall at higher $k$.

\subsubsection{Effectiveness of sentence transformer fine-tuning}\label{subsubsec:st_effectiveness}

Fine-tuning the sentence transformer improves over the pre-trained model in both settings (with and without structured filters). \emph{Without} structured filters (top table), fine-tuning with prepended metadata improves precision from 0.11 to 0.14 at $k{=}1$ and recall from 0.26 to 0.36 at $k{=}10$. The variant without prepended metadata performs best at larger $k$, achieving the highest recall at $k{=}10$ (0.36) and the highest precision at $k{=}10$ (0.08). \emph{With} structured filters (bottom table), both fine-tuned variants achieve the strongest results overall. The variant without prepended metadata attains the best $k{=}1$ scores (precision 0.32, recall 0.16) and the tied-best recall at $k{=}10$ (0.57), while the prepended-metadata variant remains very close (precision 0.30, recall 0.16 at $k{=}1$; recall 0.57 at $k{=}10$) and achieves the best or tied-best recall. Overall, fine-tuning yields consistent gains over the pre-trained configuration, and combining fine-tuning with structured filter extraction delivers the best precision and recall across all reported $k$.

\subsubsection{Does constraint-aware fine-tuning help?}
Across our baselines, fine-tuning the sentence transformer specifically to encode price, rating, and review constraints yields limited gains relative to a constraint-agnostic fine-tuning objective. Without structured filters (top table), fine-tuning improves over the pre-trained model (e.g., precision at $k{=}1$: 0.11 $\rightarrow$ 0.14; recall at $k{=}10$: 0.26 $\rightarrow$ 0.36), but the two fine-tuned variants (with vs.\ without prepended metadata) are very close and often trade places. When structured filter extraction is enabled (bottom table), 
the two fine-tuned variants remain very close at all reported $k$ (e.g., at $k{=}1$: precision 0.30 vs.\ 0.32 and recall 0.16 vs.\ 0.16; at $k{=}10$: precision 0.13 for both and recall 0.57 for both), indicating that the improvement comes from accurate pre-filtering rather than from constraint-aware representation learning.

\section{Discussion}
\noindent $\bullet$ \textbf{Effective handling of constraint information.}The experiments suggest that structured filter extraction is the primary driver of improvements in both precision and recall, as the extracted constraints effectively narrow the search space before ranking. Incorporating constraint information through metadata prepending during fine-tuning, however, does not meaningfully alter the retrieval geometry once pre-filters are applied. In practice, the effective strategy is to use a generally fine-tuned encoder for retrieval and apply structured filters afterward to enforce price, rating, and review constraints prior to ranking.
Moreover, preparing training data for constraint-aware fine-tuning is labor-intensive and less flexible. In particular, handling qualitative constraints (e.g., “cheap,” “premium”) requires ad hoc design decisions during data preparation and cannot be easily adapted post-training. By contrast, the filter extraction model converts qualitative constraints into normalized categories (high, medium, low), which can then be mapped to numerical boundaries that are adjustable at inference time. This provides a level of flexibility that is not feasible with fine-tuning alone, since any changes to constraint definitions in the fine-tuned setup would require retraining the model. Future work could explore objectives that explicitly penalize constraint violations during training, such as hard-negative mining with near-duplicate products that differ only in price or rating.

\noindent $\bullet$ \textbf{Synthetic queries generated by LLMs.} With the system effectiveness well evaluated, we now further discuss and highlight the role of the synthetic queries in our framework. Using LLMs, the synthetic queries were first generated to pair with each product and guide the fine-tuning of the sentence transformer, which is later used to compute embeddings for all products and real user queries. This process extremely contributes to the effectiveness of embedding, indicated by the performance gains described in Section~\ref{subsubsec:st_effectiveness}. Additionally, still using LLMs, the synthetic queries were enriched with rich attributes that align with realistic needs in user searches, which were later converted to structured filters of numerical and categorical information. The enriched queries and the corresponding structured filters were used to guide the fine-tuning of the user intent extraction model, which was used to accurately generate structured filters from user queries. See the performance gains described in Section~\ref{subsubsec:structure_effectiveness}. 
Results suggest that leveraging LLMs to generate and enrich synthetic queries not only overcomes the limitations of manual annotation in large datasets like ours but also effectively harnesses the world knowledge and reasoning capabilities of LLMs. Furthermore, synthetic query generation can be used to address the challenge of edge cases in search systems, for example, by instructing an LLM to generate representative edge-case queries for each product to support model training or fine-tuning.

\noindent $\bullet$ \textbf{Model efficiency.}
Our framework is designed with model efficiency in mind to reflect the real-time demands of practical e-commerce applications. Specifically, product embeddings are indexed using FAISS to enable fast similarity-based search, and a combination of lightweight models (i.e., \texttt{multi-qa-MiniLM-L6-cos-v1} and \texttt{Flan-T5-small}) is selected to handle user queries. 

\noindent $\bullet$ \textbf{Architectural support for generalization.}
Our framework is designed with flexibility in mind as well, making it well-suited for future improvements. For example, a larger embedding model can be fine-tuned and deployed for product categories where smaller models underperform. The two-stage design supports a multi-model deployment strategy across multiple product categories, where embeddings are trained separately for each category and the user intent extraction component dynamically routes queries to the most appropriate category.

\noindent $\bullet$ \textbf{Imperfect ground-truth labels.} Reviewing the retrieval results suggests that the reported precision and recall may not fully reflect the true performance of the evaluated models, as we identified labeling inconsistencies in the ESCI dataset. Several retrieved products were exact matches both semantically and in metadata but were not labeled as such in the annotations. This behavior is expected in large-scale e-commerce datasets, where relevance judgments are often incomplete or inconsistent, highlighting the challenge of evaluating real-world retrieval systems using imperfect ground-truth labels.

\noindent $\bullet$ \textbf{Limitation.} A key limitation of our work lies in the quality of the labels used to train the filter extraction model, which require manual validation due to inconsistencies in LLM-generated outputs. Fine-tuning a language model may help address this issue in future work. Additionally, the implementation and evaluation of the proposed framework focus on the \textit{Cell Phones \& Cell Phone Accessories} category. How it performs on other categories, especially those with more noise or ambiguity, remains uncertain. Nevertheless, our work demonstrates the effectiveness of a domain-specific, fine-tuned embedding model (using synthetic queries) and structured filters in capturing users’ explicit and implicit intent, providing a solid foundation for solutions in more challenging domains.

\section{Conclusion}

In this paper, we presented an LLM-based framework capable of performing effective semantic search for conversational user queries in e-commerce. Using synthetic data generated by LLMs, we fine-tuned a Sentence Transformer for product and query embedding, as well as a sequence-to-sequence model for structured filter extraction. Experiments on a test set consisting of 151 manually crafted queries and a product catalog of roughly 22 thousand items demonstrate the effectiveness and superiority of our framework across queries with varying levels of complexity, compared to baseline methods. Future work includes fine-tuning language models to improve label generation for the training task, incorporating reranking models on top of the current pipeline, and validating the viability of the approach across other product categories.

\appendix
\section{Analysis of Synthetic Queries}
We randomly sample 100 unique products (50 cell phones and 50 accessories) from the 6.5 million query–product pairs and manually screen the quality of the corresponding queries. Overall, the queries range from concise, name-based lookups to attribute-focused, conversational requests; they are relevant, diverse, and natural-sounding, and closely resemble real user behavior, with rare exceptions. We provide a few representative examples below. 

\vspace{6pt}

\textbf{Example 1:} 

\underline{Product}:
\textit{Samsung Galaxy A01 (LTE Verizon).}

\underline{Product description (excerpt)}: \textit{Long lasting battery, versatile camera, 5.7\(\,\)inch HD+ screen, expandable 16 GB storage, Android OS, Verizon compatibility.}

\underline{Generated queries}:
\begin{itemize}
  \item \textit{samsung galaxy a01 verizon 16gb}
  \item \textit{unlocked samsung galaxy a01 black}
  \item \textit{5.7 inch display phone for verizon}
  \item \textit{i need a phone with a long lasting battery}
  \item \textit{find a phone with a good camera and expandable storage}
  \item \textit{best budget phone for verizon with a large display}
  \item \textit{samsung galaxy a01 with 5.7" screen}
\end{itemize}

\underline{Observation}: 
The model mixes feature cues (battery, display, storage, and camera) with user intent (e.g., ``unlocked'' and ``long lasting battery'') and applies general knowledge about Samsung’s budget line by identifying the product as a budget phone.

\vspace{6pt}

\textbf{Example 2:}

\underline{Product}:
\textit{Temdan for Samsung Galaxy Note 20 Ultra case.}

\underline{Product description (excerpt)}: 
\textit{IP68 waterproof, built-in screen protector, full-body shockproof protection, compatible with Note 20 Ultra 5G.}

\underline{Generated queries}:
\begin{itemize}
  \item \textit{samsung galaxy note 20 ultra waterproof case}
  \item \textit{note 20 ultra case with screen protector}
  \item \textit{note 20 ultra case 360 degree protection}
  \item \textit{note 20 ultra case heavy duty}
  \item \textit{temdan note 20 ultra case}
\end{itemize}

\underline{Observation}: 
The queries vary meaningfully while consistently targeting salient attributes (IP rating, durability, and brand or model compatibility), and they reflect realistic shopping phrasing.

\vspace{6pt}
\textbf{Example 3:}

\underline{Product}:
\textit{Biter Galaxy S6/S6 Edge Case.}

\underline{Product description (excerpt):} \textit{Scratch resistant, three-in-one slim cover with electroplated frame, full-body protection; compatible with multiple Galaxy models.}

\underline{Generated queries}:
\begin{itemize}
  \item \textit{i need a galaxy s6/s6 edge case that is scratch resistant}
  \item \textit{find a galaxy s6/s6 edge case that offers full body protection}
  \item \textit{gold s6/s6 edge case}
  \item \textit{galaxy s6/s6 edge case with electroplated frame}
  \item \textit{galaxy s6/s6 edge case with three in one design}
\end{itemize}

\underline{Observation}: 
This example illustrates a rare failure case in which the generated queries exhibit limited diversity. All queries repeat the exact phrase ``s6/s6 edge,'' rather than producing distinct queries for each compatible device. In particular, no queries explicitly target ``galaxy s6'' or ``galaxy s6 edge'' individually, which reduces both query diversity and realism. In practice, users are more likely to search using a single model name rather than a combined string. This limitation could likely be mitigated by refining the prompts provided to the LLM to encourage model-specific phrasing.

That said, this behavior is not consistent across products. In several instances, the LLM correctly generated distinct queries for products compatible with multiple devices. For example, for a case compatible with both the iPhone 6 Plus and iPhone 6s Plus, the generated queries included:
\begin{itemize}
  \item \textit{iphone 6s plus case}
  \item \textit{iphone 6 plus case}
  \item \textit{iphone 6 plus / 6s plus case with mint color}
\end{itemize}

Similarly, for a Samsung battery compatible with multiple models (e.g., SGH-T919 Behold, SGH-A797 Flight), the LLM produced model-specific queries such as:
\begin{itemize}
  \item \textit{samsung battery for sgh-t919 behold}
  \item \textit{samsung battery for sgh-a797 flight}
  \item \textit{replacement battery for samsung gravity 2}
  \item \textit{i need a battery for my samsung impression}
  \item \textit{looking for a compatible battery for my samsung sgh-a727}
\end{itemize}

\bibliographystyle{ACM-Reference-Format}
\bibliography{sample-base}

@String{Computing = "Computing" }

@String{Computer = "{IEEE} Computer" }

@String{Springer = "Springer-Verlag" }

@InProceedings{Nigam19,
  author    = "Priyanka Nigam and Yiwei Song and Vijai Mohan and Vihan Lakshman and Weitian Ding and Ankit Shingavi and Choon Hui Teo and Hao Gu and Bing Yin",
  title     = "Semantic Product Search",
  booktitle = "Proceedings of the 25th ACM SIGKDD International Conference on Knowledge Discovery \& Data Mining (KDD ’19)",
  year      = "2019",
  month     = aug,
  pages     = "2876--2885",
  doi       = "10.1145/3292500.3330759",
  url       = "https://doi.org/10.1145/3292500.3330759",
}

@inproceedings{cho-etal-2022-query,
    title = "Query Generation with External Knowledge for Dense Retrieval",
    author = "Cho, Sukmin  and
      Jeong, Soyeong  and
      Yang, Wonsuk  and
      Park, Jong",
    editor = "Agirre, Eneko  and
      Apidianaki, Marianna  and
      Vuli{\'c}, Ivan",
    booktitle = "Proceedings of Deep Learning Inside Out (DeeLIO 2022): The 3rd Workshop on Knowledge Extraction and Integration for Deep Learning Architectures",
    month = may,
    year = "2022",
    address = "Dublin, Ireland and Online",
    publisher = "Association for Computational Linguistics",
    url = "https://aclanthology.org/2022.deelio-1.3/",
    doi = "10.18653/v1/2022.deelio-1.3",
    pages = "22--32"
}

@Article{Chaudhary23,
  author       = "Aditi Chaudhary and Karthik Raman and Krishna Srinivasan and Kazuma Hashimoto and Mike Bendersky and Marc Najork",
  title        = "Exploring the Viability of Synthetic Query Generation for Relevance Prediction",
  journal      = "arXiv preprint arXiv:2305.11944",
  year         = "2023",
  doi          = "10.48550/arxiv.2305.11944",
  url          = "https://doi.org/10.48550/arxiv.2305.11944",
}

@Article{Jagatap2024,
 author = {Akshay Jagatap and Srujana Merugu and Prakash Mandayam Comar},
 title = {Improving search for new product categories via synthetic query generation strategies},
 year = {2024},
 url = {https://www.amazon.science/publications/improving-search-for-new-product-categories-via-synthetic-query-generation-strategies},
}

@inproceedings{loughnane-etal-2024-explicit,
    title = "Explicit Attribute Extraction in e-Commerce Search",
    author = "Loughnane, Robyn  and
      Liu, Jiaxin  and
      Chen, Zhilin  and
      Wang, Zhiqi  and
      Giroux, Joseph  and
      Du, Tianchuan  and
      Schroeder, Benjamin  and
      Sun, Weiyi",
    editor = "Malmasi, Shervin  and
      Fetahu, Besnik  and
      Ueffing, Nicola  and
      Rokhlenko, Oleg  and
      Agichtein, Eugene  and
      Guy, Ido",
    booktitle = "Proceedings of the Seventh Workshop on e-Commerce and NLP @ LREC-COLING 2024",
    month = may,
    year = "2024",
    address = "Torino, Italia",
    publisher = "ELRA and ICCL",
    url = "https://aclanthology.org/2024.ecnlp-1.13/",
    pages = "125--135"
}

@online{LangChainSelfQuery,
  author       = "{LangChain Team}",
  title        = "{How to do “self-querying” retrieval}",
  year         = "n.d.",
  url          = "https://python.langchain.com/docs/how_to/self_query/",
  note         = "Accessed: May 22, 2025",
}

@article{douze2024faiss,
  title         = {The Faiss library},
  author        = {Matthijs Douze and Alexandr Guzhva and Chengqi Deng and Jeff Johnson and Gergely Szilvasy and Pierre-Emmanuel Mazaré and Maria Lomeli and Lucas Hosseini and Hervé J{\'e}gou},
  journal       = {arXiv preprint arXiv:2401.08281},
  year          = {2024},
  eprint        = {2401.08281},
  archivePrefix = {arXiv},
  primaryClass  = {cs.LG},
}

@online{Kashyap23,
  author    = "Karthik Kashyap",
  title     = "Consumers exit online portals due to poor search experience",
  year      = "2023",
  month     = mar,
  day       = "8",
  publisher = "Spiceworks Inc.",
  url       = "https://www.spiceworks.com/marketing/ecommerce/articles/consumers-exit-ecommerce-portal-poor-search-experience/"
}

@INPROCEEDINGS{Aamir2024,
  author={Aamir, Fatima and Sherafgan, Raheimeen and Arbab, Tehreem and Jamil, Akhtar and Bhatti, Fazeel Nadeem and Hameed, Alaa Ali},
  booktitle={2024 IEEE 3rd International Conference on Computing and Machine Intelligence (ICMI)}, 
  title={Deep Learning-based Semantic Search Techniques for Enhancing Product Matching in E-commerce}, 
  year={2024},
  volume={},
  number={},
  pages={1-9},
  doi={10.1109/ICMI60790.2024.10586148}}

@Article{Henderson17,
  author       = "Matthew L. Henderson and Rami Al-Rfou and Brian Strope and Yun-Hsuan Sung and L{\'a}szl{\'o} Luk{\'a}cs and Ruiqi Guo and Sanjiv Kumar and Balint Miklos and Ray Kurzweil",
  title        = "Efficient Natural Language Response Suggestion for Smart Reply",
  journal      = "arXiv preprint arXiv:1705.00652",
  year         = "2017",
  doi          = "10.48550/arxiv.1705.00652",
  url          = "https://doi.org/10.48550/arxiv.1705.00652",
}

@online{multiqaMinilm2024,
  author       = "{Sentence-Transformers}",
  title        = "{multi-qa-MiniLM-L6-cos-v1}",
  year         = "2024",
  organization = "Hugging Face",
  url          = "https://huggingface.co/sentence-transformers/multi-qa-MiniLM-L6-cos-v1",
  note         = "Accessed: May 22, 2025",
}

@online{SbertLosses,
  author  = "{Sentence-Transformers Team}",
  title   = "{Losses --- Sentence Transformers Documentation}",
  year    = "n.d.",
  url     = "https://sbert.net/docs/package_reference/sentence_transformer/losses.html\#multiplenegativesrankingloss",
  note    = "Accessed: May 22, 2025",
}

@inproceedings{reimers-2019-sentence-bert,
  title = "Sentence-BERT: Sentence Embeddings using Siamese BERT-Networks",
  author = "Reimers, Nils and Gurevych, Iryna",
  booktitle = "Proceedings of the 2019 Conference on Empirical Methods in Natural Language Processing",
  month = "11",
  year = "2019",
  publisher = "Association for Computational Linguistics",
  url = "https://arxiv.org/abs/1908.10084",
}

@misc{flan-t5-small,
  doi = {10.48550/ARXIV.2210.11416},
  
  url = {https://arxiv.org/abs/2210.11416},
  
  author = {Chung, Hyung Won and Hou, Le and Longpre, Shayne and Zoph, Barret and Tay, Yi and Fedus, William and Li, Eric and Wang, Xuezhi and Dehghani, Mostafa and Brahma, Siddhartha and Webson, Albert and Gu, Shixiang Shane and Dai, Zhuyun and Suzgun, Mirac and Chen, Xinyun and Chowdhery, Aakanksha and Narang, Sharan and Mishra, Gaurav and Yu, Adams and Zhao, Vincent and Huang, Yanping and Dai, Andrew and Yu, Hongkun and Petrov, Slav and Chi, Ed H. and Dean, Jeff and Devlin, Jacob and Roberts, Adam and Zhou, Denny and Le, Quoc V. and Wei, Jason},
  
  title = {Scaling Instruction-Finetuned Language Models},
  
  publisher = {arXiv},
  
  year = {2022},
  
  copyright = {Creative Commons Attribution 4.0 International}
}

@online{HuggingFaceTrainer,
  author  = "{Hugging Face}",
  title   = "{Trainer} — Transformers Documentation",
  year    = "n.d.",
  url     = "https://huggingface.co/docs/transformers/en/main_classes/trainer?utm_source=chatgpt.com\#transformers.Seq2SeqTrainer",
  note    = "Accessed: May 22, 2025",
}

@article{DBLP:journals/corr/abs-1810-04805,
  author    = {Jacob Devlin and
               Ming{-}Wei Chang and
               Kenton Lee and
               Kristina Toutanova},
  title     = {{BERT:} Pre-training of Deep Bidirectional Transformers for Language
               Understanding},
  journal   = {CoRR},
  volume    = {abs/1810.04805},
  year      = {2018},
  url       = {http://arxiv.org/abs/1810.04805},
  archivePrefix = {arXiv},
  eprint    = {1810.04805},
  bibsource = {dblp computer science bibliography, https://dblp.org}
}

@book{jurafsky2025slp3,
  author    = {Daniel Jurafsky and James H. Martin},
  title     = {Speech and Language Processing: An Introduction to Natural Language Processing, Computational Linguistics, and Speech Recognition with Language Models},
  edition   = {3rd},
  year      = {2025},
  note      = {Online manuscript released January 12, 2025},
  url       = {https://web.stanford.edu/~jurafsky/slp3}
}

@article{hou2024bridging,
  title={Bridging Language and Items for Retrieval and Recommendation},
  author={Hou, Yupeng and Li, Jiacheng and He, Zhankui and Yan, An and Chen, Xiusi and McAuley, Julian},
  journal={arXiv preprint arXiv:2403.03952},
  year={2024}
}

@misc{google_gemini_api_models_2025,
  title        = {Gemini Models},
  author       = {{Google AI for Developers}},
  year         = {2025},
  month        = {May},
  url          = {https://ai.google.dev/gemini-api/docs/models},
  note         = {Accessed: 2025-05-24}
}

@misc{wallace2019do,
  title        = {Do NLP Models Know Numbers? Probing Numeracy in Embeddings},
  author       = {Wallace, Eric and Wang, Yizhong and Li, Sujian and Singh, Sameer and Gardner, Matt},
  year         = {2019},
  howpublished = {arXiv preprint arXiv:1909.07940},
  doi          = {10.48550/arxiv.1909.07940},
  url          = {https://doi.org/10.48550/arxiv.1909.07940}
}

@article{kim-etal-2019-categorical,
    title = "Categorical Metadata Representation for Customized Text Classification",
    author = "Kim, Jihyeok  and
      Amplayo, Reinald Kim  and
      Lee, Kyungjae  and
      Sung, Sua  and
      Seo, Minji  and
      Hwang, Seung-won",
    editor = "Lee, Lillian  and
      Johnson, Mark  and
      Roark, Brian  and
      Nenkova, Ani",
    journal = "Transactions of the Association for Computational Linguistics",
    volume = "7",
    year = "2019",
    address = "Cambridge, MA",
    publisher = "MIT Press",
    url = "https://aclanthology.org/Q19-1013/",
    doi = "10.1162/tacl_a_00263",
    pages = "201--215"
}

@article{vaswani2017attention,
  title={Attention is all you need},
  author={Vaswani, Ashish and Shazeer, Noam and Parmar, Niki and Uszkoreit, Jakob and Jones, Llion and Gomez, Aidan N and Kaiser, {\L}ukasz and Polosukhin, Illia},
  journal={Advances in neural information processing systems},
  volume={30},
  year={2017}
}

@inproceedings{reimers2019sentence,
  title={Sentence-BERT: Sentence Embeddings using Siamese BERT-Networks},
  author={Reimers, Nils and Gurevych, Iryna},
  booktitle={Proceedings of the 2019 Conference on Empirical Methods in Natural Language Processing and the 9th International Joint Conference on Natural Language Processing (EMNLP-IJCNLP)},
  pages={3982--3992},
  year={2019}
}

@inproceedings{radlinski2017theoretical,
  title={A theoretical framework for conversational search},
  author={Radlinski, Filip and Craswell, Nick},
  booktitle={Proceedings of the 2017 conference on conference human information interaction and retrieval},
  pages={117--126},
  year={2017}
}

@inproceedings{liu2020enhancing,
  title={Enhancing multi-turn dialogue modeling with intent information for E-commerce customer service},
  author={Liu, Ruixue and Chen, Meng and Liu, Hang and Shen, Lei and Song, Yang and He, Xiaodong},
  booktitle={Natural Language Processing and Chinese Computing: 9th CCF International Conference, NLPCC 2020, Zhengzhou, China, October 14--18, 2020, Proceedings, Part I 9},
  pages={65--77},
  year={2020},
  organization={Springer}
}

@inproceedings{yang2018query,
  title={Query tracking for e-commerce conversational search: A machine comprehension perspective},
  author={Yang, Yunlun and Gong, Yu and Chen, Xi},
  booktitle={Proceedings of the 27th ACM International Conference on Information and Knowledge Management},
  pages={1755--1758},
  year={2018}
}

@inproceedings{liu2023u,
  title={U-need: A fine-grained dataset for user needs-centric e-commerce conversational recommendation},
  author={Liu, Yuanxing and Zhang, Weinan and Dong, Baohua and Fan, Yan and Wang, Hang and Feng, Fan and Chen, Yifan and Zhuang, Ziyu and Cui, Hengbin and Li, Yongbin and others},
  booktitle={Proceedings of the 46th international ACM SIGIR conference on research and development in information retrieval},
  pages={2723--2732},
  year={2023}
}

@inproceedings{jia2022convrec,
  title={E-ConvRec: A large-scale conversational recommendation dataset for E-commerce customer service},
  author={Jia, Meihuizi and Liu, Ruixue and Wang, Peiying and Song, Yang and Xi, Zexi and Li, Haobin and Shen, Xin and Chen, Meng and Pang, Jinhui and He, Xiaodong},
  booktitle={Proceedings of the Thirteenth Language Resources and Evaluation Conference},
  pages={5787--5796},
  year={2022}
}

@article{reddy2022shopping,
title={Shopping Queries Dataset: A Large-Scale {ESCI} Benchmark for Improving Product Search},
author={Chandan K. Reddy and Lluís Màrquez and Fran Valero and Nikhil Rao and Hugo Zaragoza and Sambaran Bandyopadhyay and Arnab Biswas and Anlu Xing and Karthik Subbian},
year={2022},
eprint={2206.06588},
archivePrefix={arXiv}
}

@misc{faiss_wiki_indexes,
  title        = {FAISS Wiki: Faiss indexes},
  author       = {{FAISS Team}},
  howpublished = {\url{https://github.com/facebookresearch/faiss/wiki/Faiss-indexes}},
  note         = {Accessed: 2025-10-08}
}

@misc{Reid2025AIModeSearch,
  author       = {Elizabeth Reid},
  title        = {{AI} Mode in Google Search: Updates from Google {I/O} 2025},
  year         = {2025},
  month        = may,
  howpublished = {\url{https://blog.google/products/search/google-search-ai-mode-update/\#ai-mode-search}},
  note         = {Accessed: 2025-12-22}
}

@misc{Bloomreach2025ConversationalAIShoppingStudy,
  author       = {{Bloomreach}},
  title        = {More Than 60\% of Consumers Have Used Conversational {AI} for Shopping, New Research From Bloomreach Finds},
  year         = {2025},
  month        = mar,
  howpublished = {\url{https://www.bloomreach.com/en/news/2025/bloomreach-announces-findings-from-conversational-ai-shopping-study/}},
  note         = {Accessed: 2025-12-22}
}

\end{document}